\documentclass[aps,pra,superscriptaddress,showfacs,twocolumn,showpacs]{revtex4}
\usepackage{amsmath}
\usepackage{graphicx}
\usepackage{amssymb}

\usepackage{bm,bbm, color}

\newcommand*{\cl}[1]{{\mathcal{#1}}}

\newcommand*{\tn}[1]{{\textnormal{#1}}}
\newcommand{\ket}[1]{\left|#1\right>}

\newcommand{\inn}[2]{\left<#1|#2\right>}

\newcommand*{\1}{{\mathbbm{1}}}

\begin{document}

\title{Non-Hermitian Hamiltonian and Lamb shift in circular dielectric microcavity}

\author{Kyu-Won Park}
\affiliation{Department of Physics, Sogang University, Seoul 04107, Korea}
\affiliation{School of Computational Sciences, Korea Institute for Advanced Study, Seoul 02455, Korea}
\author{Jaewan Kim}
\affiliation{School of Computational Sciences, Korea Institute for Advanced Study, Seoul 02455, Korea}
\author{Kabgyun Jeong}
\affiliation{School of Computational Sciences, Korea Institute for Advanced Study, Seoul 02455, Korea}
\affiliation{Center for Macroscopic Quantum Control, Department of Physics and Astronomy, \\
Seoul National University, Seoul 08826, Korea}

\date{\today}

\begin{abstract}
We study the normal modes and quasi-normal modes (QNMs) in circular dielectric microcavities through non-Hermitian Hamiltonian, which come from the modifications due to system-environment coupling. Differences between the two types of modes are studied in detail, including the existence of resonances tails. Numerical calculations of the eigenvalues reveal the Lamb shift in the microcavity due to its interaction with the environment. We also investigate relations between the Lamb shift and quantized angular momentum of the whispering gallery mode as well as the refractive index of the microcavity. For the latter, we make use of the similarity between the Helmholtz equation and the Schr\"{o}dinger equation, in which the refractive index can be treated as a control parameter of effective potential. Our result can be generalized to other open quantum systems with a potential term.

\end{abstract}
\maketitle

\section{Introduction}
The so-called quantum billiard systems are conservative closed systems with Dirichlet boundary condition described by Hermitian Hamiltonian with real eigenvalues~\cite{KKS99,S99}. Once these systems become open, i.e., coupled to their environment, the situation changes. Generally these systems are described by a non-Hermitian Hamiltonian with complex eigenvalues~\cite{R09,D00,R13}. One formalism for treating open systems was developed by Feshbach in 1958 to deal with nuclear decay~\cite{F58}. Since then the formalism has been used in many other fields such as atomic physics~\cite{M11}, solid state physics~\cite{CK09}, microwave cavities~\cite{PRSB00}. 

The Feshbach projection operator formalism yields non-Hermitian Hamiltonian, describing various kinds of interesting phenomena such as bi-orthogonality~\cite{L09}, phase rigidity~\cite{BRS07}, avoided resonance crossing~\cite{RLK09,SGLX14,W06}, and exceptional points~\cite{LM+09,OP09,MR08}. One prominent example is the Lamb shift. It describes a small energy shift in a quantum system caused by the vacuum fluctuations~\cite{LR47,SS10,BN+94}. Initially it was observed for a hydrogen atom, but recently the effect has been studied in photonic crystals~\cite{WKG04} and cavity QED systems~\cite{N10}. Here, we generalize this to dielectric microcavities.

As a microcavity is a very attractive optical source in optoelectronic circuits, their emission patterns and high quality factors have been investigated in detail both experimentally and theoretically~\cite{GC+98,NS97,KPYK11,SF+09}. Importantly, they provide a good platform~\cite{CW15,W06} to study  the above-mentioned phenomena in open-quantum systems such as bi-orthogonality, phase rigidity, avoided resonance crossing, and exceptional points. We thus investigate the Lamb shift in these systems by comparing the real eigenvalues of the circular quantum billiards and dielectric microwave cavities. Investigations on the openness effects are essential, as exemplified by the concepts such as quasi-scar~\cite{LR+04} and Goos-H\"{a}nchen shift~\cite{SH06,SG+10}.
 
This article is organized as follows. In Sec.~\ref{Helm} we briefly review the theoretical descriptions of quantum billiards and dielectric cavities. This is followed by a summary of Feshbach projection operator (FPO) formalism in Sec.~\ref{GHeff}, which yields non-Hermitian Hamiltonians. The resulting quasi-normal modes are compared to the normal modes of quantum billiards in Sec.~\ref{QNM}. Section~\ref{LCC} presents our main results on Lamb shift in a single-layer whispering gallery mode (sec~\ref{WGM}), and Lamb shift as a function of the refractive index $n$ (sec~\ref{refractive}). We summarise our results and conclude in Sec.~\ref{conclusion}.

\section{Quantum billiard and dielectric microcavity}\label{Helm}
While quantum billiards are completely closed system, the dielectric microcavity is an open quantum system. The eigenvalues of (time-independent) Hamiltonian of a quantum billiard system and those of dielectric microcavity are solutions of Helmholtz equation with different boundary conditions~\cite{J75} which can be solved by boundary element methods~\cite{W03}.
In quantum billiards, the wave function $\psi$ is described by the Helmholtz equation
\begin{equation} \label{eq:helm}
(\nabla^2+k^2)\psi(\bm{r})=0,
\end{equation}
where $k$ is wave number and $\psi(\bm{r})$ is a wave function such that $\bm{r}$ is a position vector inside the billiard system. Its boundary conditions are given by
\begin{align*}
\psi(\bm{r}=R)&=0,~~~~(\textnormal{Dirichlet})  \\
\partial_n\psi(\bm{r}=R)&=0,~~~~(\textnormal{Neumann}) \\
\end{align*}
where $R$ indicates the boundary of the system, $\partial_n$ is the normal derivative to the boundary. Note that we always fix $R=1$ in this paper. The eigenvalues are always real for quantum billiards defined by the above conditions. For a dielectric microcavity with a refractive index $n$, $k^2$ of Eq.~(\ref{eq:helm}) must be replaced by $n^2k^2$. In this work, we set $n$ to be 3.3, which is the refractive index of InGaAsP semiconductor microcavity~\cite{KPYK11}.
Therefore, the TE mode $\psi(\bm{r})$ obeys the equation
\begin{align}
\nabla^2\psi(\bm{r})+k^2\mu(\bm{r})\psi(\bm{r})=0,
\end{align}
where $\mu(\bm{r})=1+(n^2-1)\Theta(R-\bm{r})$, $n$ is the refractive index of the cavity, $\Theta(\cdot)$ is the unit step function, and its boundary conditions are given by
\begin{align*}
\psi_{in}(\bm{r}=R)&=\psi_{out}(\bm{r}=R), \\
\partial_n\psi_{in}(\bm{r}=R)&=n^2\partial_n\psi_{out}(\bm{r}=R).
\end{align*}
In this paper, we consider only the TE modes. Especially, the solutions of a circular cavity are Bessel functions, classified by a radial quantum number $\ell$ and angular quantum number $m$.

\section{Non-Hermitian Hamiltonian description of open quantum system} \label{GHeff}
Typically, quantum systems are not completely closed, i.e., obey the boundary condition above Eq.~(\ref{eq:helm}), but are coupled to their environment. In this case, the total Hilbert space consists of the quantum billiard $S$ and an environment $E$. The total system obeys the Schr\"{o}dinger equation
\begin{align} \label{eq:schrodinger}
\bm{H}\ket{\cl{E}}_{SE}=\cl{E}\ket{\cl{E}}_{SE},
\end{align}
where $\bm{H}$ is total Hamiltonian with real energy eigenvalues, and $\cl{E}$ is the total energy of the Hamiltonian with the corresponding eigenstate $\ket{\cl{E}}_{SE}$. Note that $\ket{\cl{E}}$ is the corresponding eigenstates of the energy $\cl{E}$.
The quantum billiard has a discrete set of states, while we assume that the environment has a continuous set. We can define the projection operators $\Pi_S$ and $\Pi_E$, with $\Pi_S+\Pi_E=\1_{SE}$ and $\Pi_S\Pi_E=\Pi_E\Pi_S=0$. Here, $\Pi_S$ is a projection onto the quantum billiard system whereas $\Pi_E$ is a projection onto the environment, and $\1_{SE}$ is a identity operator defined on the total space~\cite{R09,D00}.
In this case, the total Hamiltonian is generally given by
\begin{equation}
\bm{H}=H_{S}+H_{E}+V_{SE}+V_{ES},
\end{equation}
where $H_{S}=\Pi_S\bm{H}\Pi_S$ and $H_{E}=\Pi_E\bm{H}\Pi_E$ are the Hamiltonian of the quantum billiard system and environment, respectively, and $V_{SE}=\Pi_S\bm{H}\Pi_E\equiv V$ and $V_{ES}=\Pi_E\bm{H}\Pi_S\equiv V^\dagger$ are interaction Hamiltonians between the system and the environment: $V_{SE}$ is the interaction from $E$ to $S$ and $V_{ES}$ vice versa.

By exploiting the Hamiltonian~\cite{R09,D00,R13},
we can derive an \emph{effective} non-Hermitian Hamiltonian defined solely on quantum billiard system $S$:
\begin{align}
H_{\textnormal{eff}}
&=H_{S}+V_{SE}G_E^+V_{ES} \label{eq:effham1}\\
&=H_{S}-\frac{1}{2}iVV^\dagger+P\int_{a}^b d\cl{E}' \frac{VV^\dagger}{\cl{E}-\cl{E}'}. \label{eq:effham2}
\end{align}
Here, $G_E^+$ is an energy dependent out-going Green function, $VV^\dagger$ is the system interaction via the environment, and $P$ means the principal value. The integral domain $[a,b]$ is known as energy window determined by each decay channels. The decay channels are well-known in a few cases such as rectangular waveguides~\cite{D00}, but in dielectric microcavities, it is only known that the decay channel corresponds to each resonance modes or quasi-normal modes (QNMs) and they are characterized by far-field patterns~\cite{SGLX14}. Since there is only a single decay channel with respect to each resonance mode, we do not need a summation of decay channels for the principal value in Eq.~(\ref{eq:effham2}). Eq.~(\ref{eq:effham1}) can be transformed to Eq.~(\ref{eq:effham2}) by Sokhotski-Plemelj theorem~\cite{M98}. After all, this non-Hermitian Hamiltonian has generally complex eigenvalues instead of real eigenvalues:
\begin{equation} \label{eq:effham3}
H_{\textnormal{eff}}\ket{\phi_{j}}=z_{j}\ket{\phi_{j}},\;\;
z_j=\cl{E}_j-\frac{i}{2}\Gamma_j,
\end{equation}
where $z_{j}$ is a complex number with $\cl{E}_j$ and $\Gamma_j$ representing the energy and decay width of $j$-th eigenvector, respectively. Hence, the quality factor $Q$, as a measure of energy-storing capability, is defined by $Q=\frac{\cl{E}_j}{2\Gamma_j}$. In addition, the eigenvectors $\ket{\phi_j}$'s are generally not orthogonal but bi-orthogonal states satisfying $\inn{\phi_{i}^L}{\phi_{j}^R}=\delta_{ij}$~\cite{R09}. Note that $L$ and $R$ denote left and right eigenvectors, respectively.

\section{Normal mode versus quasi-normal modes}\label{QNM}

\begin{figure}
\centering
\includegraphics[width=8.5cm]{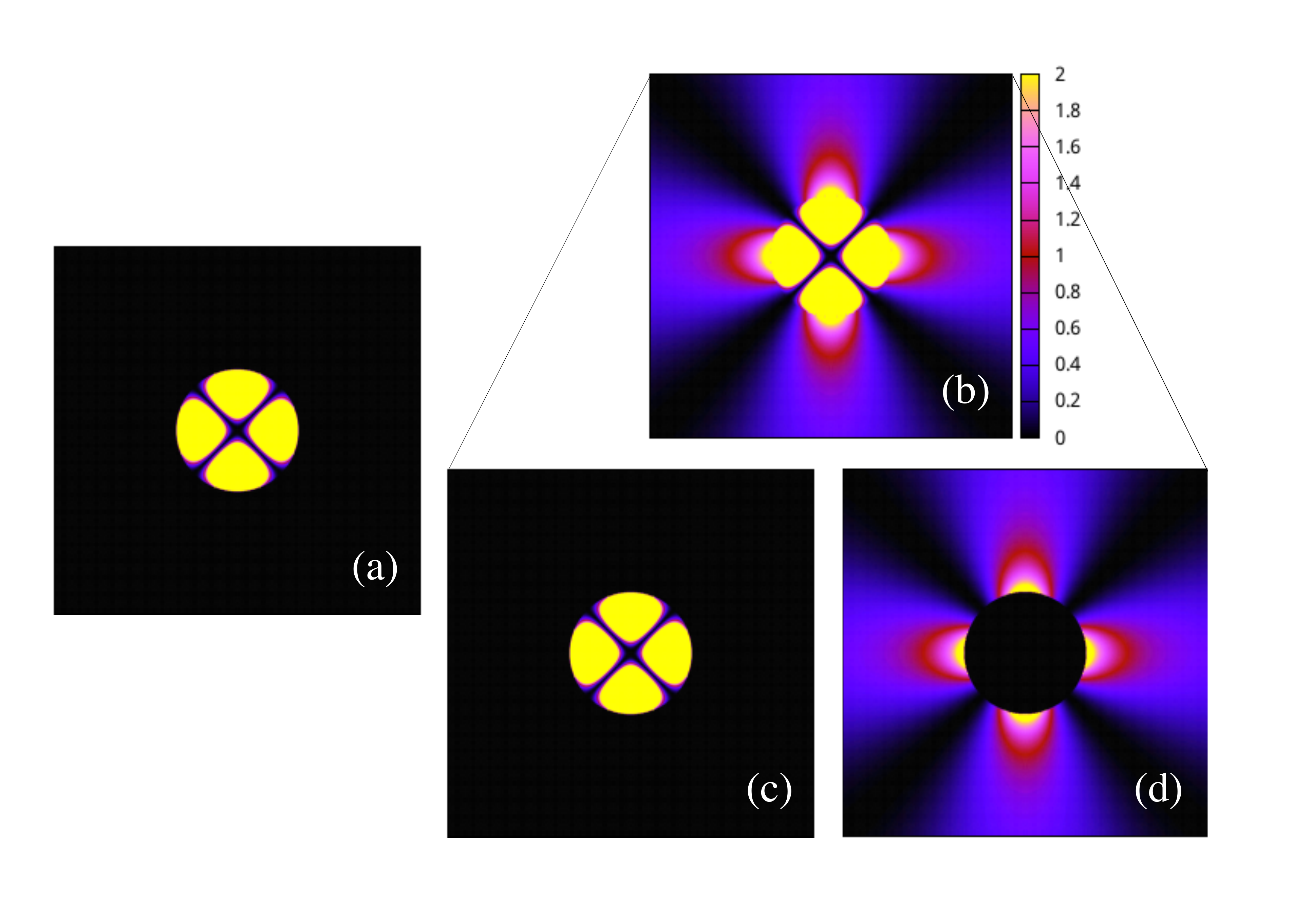}
\caption{(Color online) Intensity plot of eigenfunctions. (a) Normal mode with $\ell=1$, $m=2$, and refractive index $n=3.3$ in circular billiard. It has always real eigenvalues, i.e., $\tn{Re}(kR)\cong1.556$. (b) Quasi-normal mode (QNM) with same quantum number and refractive index in circular dielectric microcavity. It can exist out of the cavity. The QNM consists of two functions, i.e., $\ket{\phi_k}$ and $\ket{\omega_{k}}$. (c) The eigenfunction $\ket{\phi_{k}}$ of $H_{\tn{eff}}$ as an interior part of the QNM has complex eigenvalues  with $\tn{Re}(kR)\cong1.462$ and $\tn{Im}(kR)\cong-0.046$, and it is defined solely in system subspace $S$. (d) The resonance tail $\ket{\omega_{k}}$ is defined in environmental subspace $E$, and it plays role of the decay channel.}
\label{Figure-1}
\end{figure}

The normal modes are eigenfunctions of the Hermitian Hamiltonian $(H_{S})$ with real eigenvalues and are strictly localized in the interior of the system. On the other hand, the quasi-normal modes can leak out into the environment. They can be written as a sum of two parts~\cite{R09}:
\begin{align}
\ket{\Omega_j}&=\ket{\phi_j}+G_E^+V_{ES}\ket{\phi_j}\\
&\equiv\ket{\phi_j}+\ket{\omega_{j}}.
\end{align}
The $\ket{\phi_j}$ are eigenstates of $H_\tn{eff}$ with generally complex eigenvalues and are localized inside the system, and $\ket{\omega_j}$ describe the resonance tails that reside entirely in the environment. Its presence can be explained as follows.
The interaction term $V_{ES}$ gives connection between the system and the environment so that the interior eigenfunction can leak out of the system, which then propagates through environment by out-going Green function $G_E^+$. As pointed out in Sec.~\ref{GHeff}, the decay channels are  characterized by far-field patterns, thus we should consider the decay channel as a resonance tail rather than resonance itself, because it is defined only in subspace $E$. Therefore, this $\ket{\omega_{k}}$ plays the role of decay channel in dielectric microcavity.

The above description holds well for microcavities as shown in Fig.~\ref{Figure-1}. We can check the above situation from (a) and (b) in Fig.~\ref{Figure-1}. Figure ~\ref{Figure-1}(a) shows a normal mode of circular billiard quantum system with radial quantum number $\ell=1$, angular quantum $m=2$, and refractive index $n=3.3$. In this case, we found $\tn{Re}(kR)\cong 1.556$. The wave function of the quantum billiard system resides entirely with in the boundary. It can be conformed by Figure~\ref{Figure-1}(a), since the brighter points represent the higher probability. Note that the completely black region of outside of the cavity reflects this fact. 
 
 On the other hand, Fig.~\ref{Figure-1}(b) depicts a QNM with same quantum numbers ($\ell$ and $m$) and refractive index ($n$). We can easily identify that (b) is made of (c) and (d). The figure (c) depicts the eigenfunction $\ket{\phi_{k}}$ of $H_{\tn{eff}}$ as interior part of QNM with complex eigenvalue $\tn{Re}(kR)\cong1.462$ and $\tn{Im}(kR)\cong-0.0463$. Note that the real eigenvalue of circular billiard is greater than that of circular dielectric cavity. In this case, the difference between the real part of the eigenvalues of the normal mode and QNM is nearly 0.094. Figure~\ref{Figure-1}(d) depicts resonance tail $\ket{\omega_{k}}$ with completely black region inside the cavity, which means that it is completely defined on subsystem $E$.
 
\section{Lamb shift in a circular microcavity} \label{LCC}
\subsection{Whispering gallery mode and angular quantum number} \label{WGM}
The non-Hermitian Hamiltonian predicts various kinds of interesting concepts such as complex eigenvalues, bi-orthogonality, phase rigidity, Lamb shift, exceptional points and so on. Among these, the Lamb shift is a small difference in energy levels of a hydrogen atom in quantum electrodynamics caused by the vacuum fluctuations. In this paper, we consider, for the first time in our knowledge, the Lamb shift in dielectric microcavities. 
This Lamb shift is also well formalized in non-Hermitian Hamiltonian by following equations. First of all, we divide the effective Hamiltonian $H_{\tn{eff}}$ into real and imaginary parts, respectively. From Eq.~(\ref{eq:effham2}), the real and imaginary parts of $H_{\tn{eff}}$ are given by
 \begin{align}
\tn{Re}(H_{\tn{eff}})&=H_{S}+P\int_{a}^b d\cl{E}' \frac{VV^\dagger}{\cl{E}-\cl{E}'}\;\;\tn{and} \label{eq:re}\\
\tn{Im}(H_{\tn{eff}})&=-\frac{1}{2}VV^\dagger. \label{eq:im}
\end{align}
Eq.~(\ref{eq:re}) represents always real matrix, because $H_{S}$ and $VV^\dagger$ are Hermitian matrix having real eigenvalues. On the contrary, Eq.~(\ref{eq:im}) represents imaginary matrix, since the term of $VV^\dagger$ itself is a Hermitian matrix but the imaginary value $i$ is multiplied as in Eq.~(\ref{eq:effham2}), hence its eigenvalue is always a pure imaginary number. So, the small energy difference (Lamb shift) between closed and open systems~\cite{R09,R13} can be obtained by
\begin{align} \label{eq:deltaE}
L\equiv\tn{Re}(H_{\tn{eff}})-H_{S}=P\int_{a}^b d\cl{E}' \frac{VV^\dagger}{\cl{E}-\cl{E}'}.
\end{align}
Even though the specific form of the principal value integral in dielectric cavity is not yet known, we can see that the numerator of integrand in Eq.~(\ref{eq:deltaE}) is $VV^\dagger$ and it is equivalent to $\tn{Im}(kR)$.

\begin{figure}
\centering
\includegraphics[width=9.2cm]{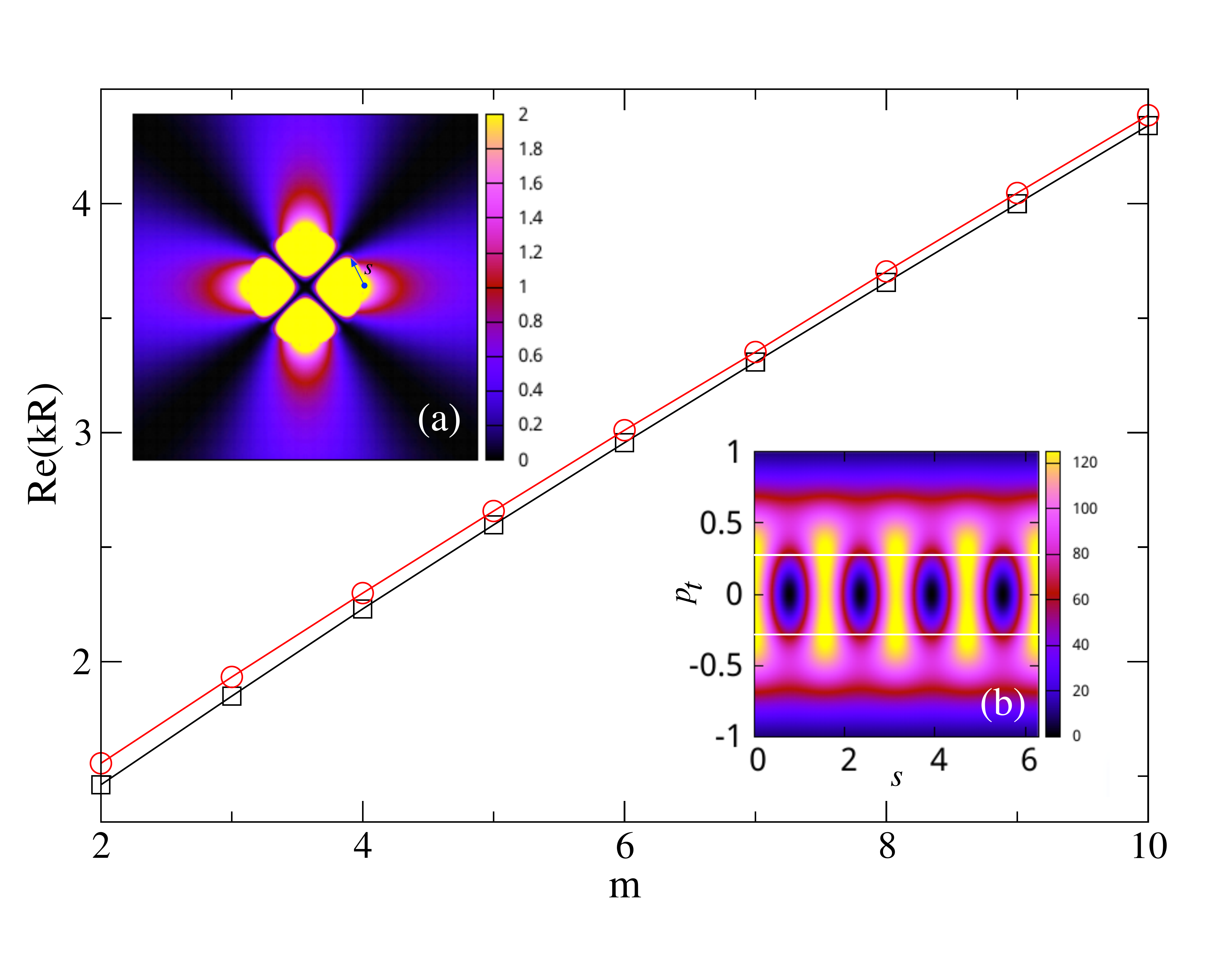}
\caption {(Color online) $\tn{Re}(kR)$ in the circular billiard and the dielectric microcavity as a function of $m$ for $\ell=1$. The red circles are eigenvalues of billiard, and black squares are those of dielectric cavity. The difference between the two modes, which corresponds to the Lamb shift $L$, gets smaller with increasing $m$. The insets shows wave intensity (a) and its Husimi function (b) with $\ell=1$ and $m= 2$. The Husimi function in the inset (b) concentrates on the upper white critical line with $p_t\simeq0.3$ for $n=3.3$. Note that $s$ is an arc length of the cavity from 0 to $2\pi$, and $p_t=\sin\chi$ is tangential momentum whose incident angle $\chi$ are defined at normal vector on each boundary point.}
\label{Figure-2}
\end{figure}

There are various kinds of resonance modes in dielectric microcavities. However, the whispering gallery modes (WGMs) are very significant one among them, because it ensures a high quality factor. So, we adopt WGMs as first example of the Lamb shift. The WGM is a resonance mode which is localized near boundary of the cavity to be confined by total internal reflection so that it can not leak out in classical regime. We restrict the modes to $\ell=1$, and $m\geq2$ to fulfill typical WGM conditions mentioned above. This is verified by inset in Fig.~\ref{Figure-2}. The inset (a) in Fig.~\ref{Figure-2} is resonance mode with $\ell=1$, $m=2$, and shows a starting point of arc length $s$. The inset (b) in Fig.~\ref{Figure-2} is its Husimi function with critical line. The $X$-axis of it is an arc length $s$ from 0 to $2\pi$ and $Y$-axis is $p_t=\sin\chi$ as tangential momentum whose incident angle $\chi$ is defined at normal vector on each boundary point. The white colored critical line is $p_t\simeq0.3$ for $n=3.3$ of InGaAsP semiconductor microcavity laser. Major part of Husimi function lies on the upper white critical line.

\begin{figure}
\centering
\includegraphics[width=9.2cm]{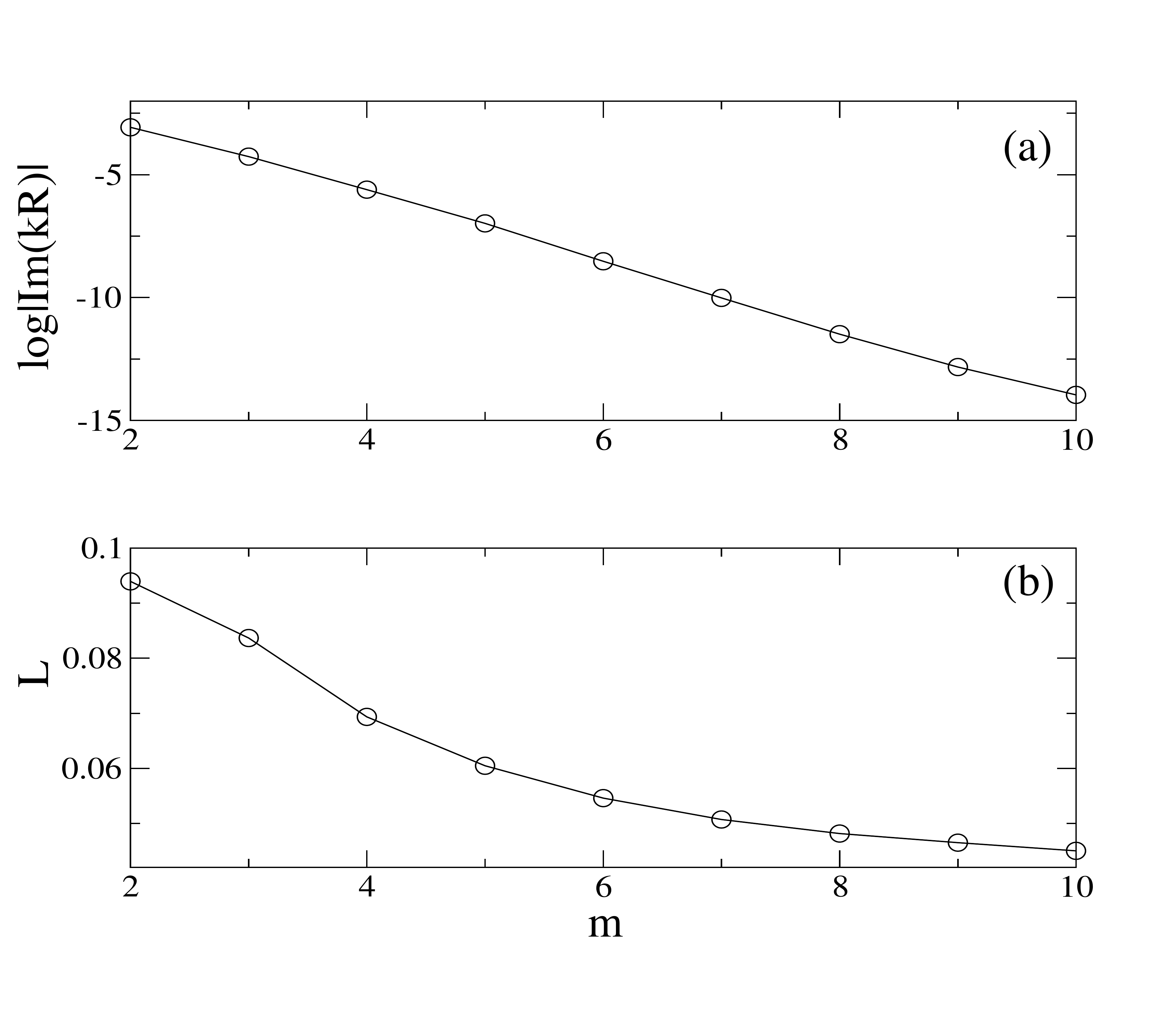}
\caption {(a) The absolute value of $\tn{Im}(kR)$ in logscale of WGM in Fig.~\ref{Figure-2} versus angular momentum $m$. These values are getting reduced such that WGM is getting way from critical angles through more adhering to boundary as $m$ increases. (b) Lamb shift of WGM in Fig.~\ref{Figure-2} versus angular momentum $m$. These value also decreases as $m$ increases.}
\label{Figure-3}
\end{figure}

There are two kinds of real eigenvalues of single layer WGM depicted in Fig.~\ref{Figure-2}. The red circles are eigenvalues of circular billiard with increasing  $\textit{m}$, while black squares are those of circular microcavity. Both of them increases almost linearly as $m$ increases, but not exactly. Careful examination shows that the difference or Lamb shift $L$ decreases as $m$ grows. Figure~\ref{Figure-3}(a) plots the absolute value of $\tn{Im}(kR)$ in logscale of WGM in Fig.~\ref{Figure-2} versus $m$, and Fig.~\ref{Figure-3}(b) is the Lamb shift $L$ versus $m$. The latter shows us that $L$ decreases rapidly in $2\leq m \leq6$ and thereafter decreases asymptotically. Figure~\ref{Figure-3}(a) shows  similar pattern to (b). It decreases almost exponentially with $m$. (Note that Figure~\ref{Figure-3}(a) is plotted in logscale.) The reason why absolute value of $\tn{Im}(kR)$ decreases as $m$ increases is obvious. As $m$ increases, the WGM is getting more adhering to boundary of cavity such that it is getting away from critical angle to have less decay. 

Regarding WGMs, as angular quantum number $m$ increases, the Lamb shift $L$ decreases since the numerator of integrand in Eq.~(\ref{eq:deltaE}) is proportional to absolute value of $\tn{Im}(kR)$. These results are in accord with our explanation.

\subsection{Refractive index $n$ as a potential well}\label{refractive}
It has been investigated the relation between refractive index $n$ and potentials~\cite{CC96,TVC09}.
In dielectric microcavities, the refractive index $n$ is tunable as a system parameter in theory. This fact also gives us powerful method to reveal the character of Lamb shift, since it implies that we can control the strength of system-environment interactions systemetically. Intuitively, as $n$ goes to infinity, the microcavity converges to a billiard quantum system. To verify this conjecture, it is convenient to take a scattering process. That is, we rewrite the Helmholtz equation to a Schr\"{o}dinger-like form by exchanging the electric or magnetic field for the wave function $\psi(\bm{r})$. Then, the Hamiltonian of circular dielectric disk~\cite{CK+10,H02} is given by
\begin{align}
\left(-\nabla^2+k^2(1-n^2)\right)\psi(\bm{r})=k^2\psi(\bm{r}),
\end{align}
where $n$ is the refractive index of the system, ${k^2}$ is a real energy $\cl{E}$, and we set the refractive index of the exterior to be 1, same as vacuum. Then if we set $\psi(r,\phi)=\psi(r)e^{im\phi}$ in polar coordinate, the angular part of solution is just $e^{im\phi}$ and the radial part of that is 
\begin{align}
-\left[\frac{d^2}{dr^2}+\frac{1}{r}\frac{d}{dr}\right]\psi(r)+V_{\tn{eff}}(r)\psi(r)=k^2\psi(r),
\end{align}
where the effective potential is given by
\begin{equation} \label{eq:effpot}
V_{\tn{eff}}(r)=k^2\left[1-n^2(r)\right]+\frac{m^2}{r^2}
\end{equation}
with angular quantum number $m$. It is important to note that the radial quantum number $\ell$ is irrelevant to $V_{\tn{eff}}(r)$. Since $n$ is typically greater than 1, the first term of $V_{\tn{eff}}$ must always be negative. Therefore it can be interpreted as an attractive potential well and the second term is a repulsive angular momentum barrier.
The classical turning points on a potential well are described by
\begin{align} \label{eq:classicalTP}
\cl{E}-V_{\tn{eff}}(r)=\left(k^2n^2-\frac{m^2}{r^2}\right)=0.
\end{align}
So, this turning point at boundary $r=R$ gives us trapped region whose top is $k^2_T$ as  $(m/R)^2$ and bottom is $k^2_B$ as $(m/nR)^2$ originated from discontinuity of $n$ at the boundary $R$. We set $R=1$ again. It is notable that $k_T$ is fixed completely by only the angular momentum $m$, not $n$ or $k^2$. Even $k_T$ does not depend on $\ell$. Thus, the eigenfunction must lie between these values to be trapped in potential well. This is known as \emph{below-barrier resonances}~\cite{CK+10,H02}, i.e.,
\begin{equation}
\frac{m}{n}<\tn{Re}(kR)< m.
\end{equation}
Furthermore, these trapped states must lead to the small $\tn{Im}(kr)$ and higher $Q$ factor of resonances, so we are certain that WGM is trapped. 

\begin{figure}
\centering
\includegraphics[width=9.2cm]{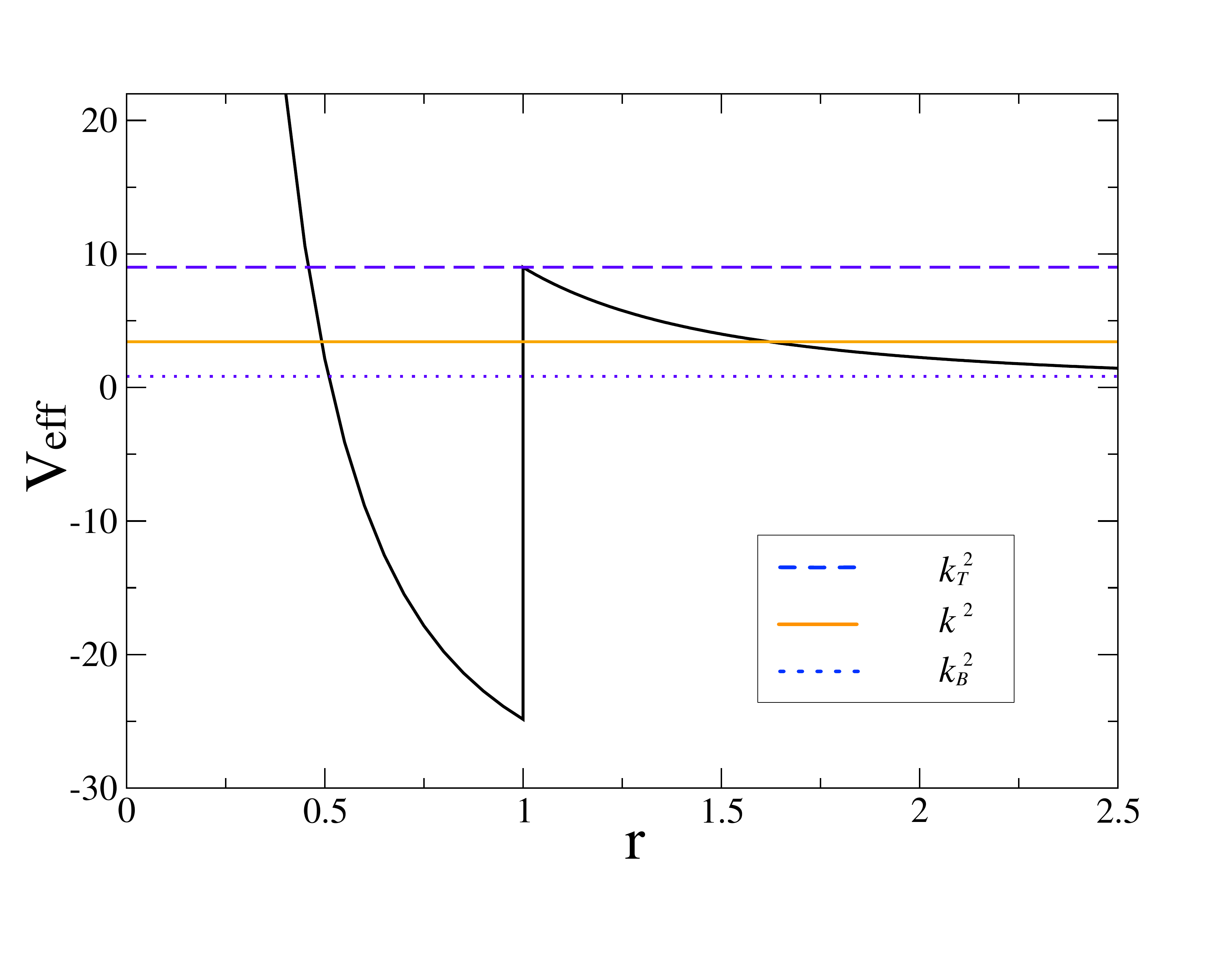}
\caption {(Color online) Effective potential for trapped resonance for $\ell=1$ and $m=3$, with $n=3.3$ and $R=1$. Blue dashed line is $k_T^2=9$, solid orange line is $\tn{Re}(kR)^2=\cl{E}\cong 3.42$, and blue dotted line is $k_B^2\cong0.826$.}
\label{Figure-4}
\end{figure}

\begin{figure}
\centering
\includegraphics[width=9.2cm]{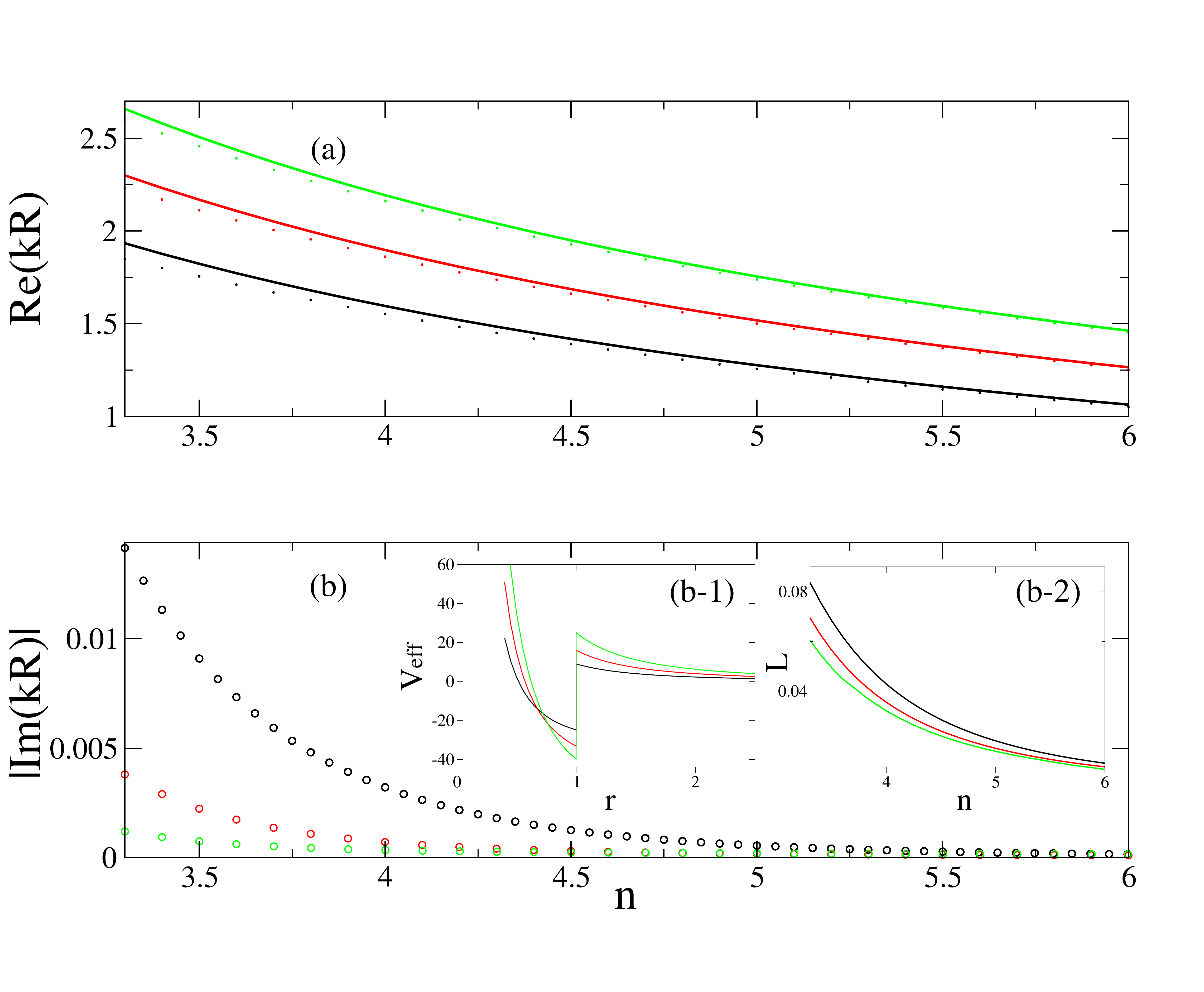}
\caption {(Color online) (a) The trace of $\tn{Re}(kR)$ in circular billiard and dielectric cavity versus increasing refractive index $n$ with fixed radial quantum number $\ell=1$ and angular quantum number $m=3,4,5$. The dotted lines are eigenvalues of dielectric cavity whereas solid lines are those of circular billiard. The black colors are for $m=3$ and red colors are for $m=4$ and green colors are for $m=5$, respectively. The relative widths of dotted and solid line are getting smaller as $n$ increases. These are well seen as Lamb shift $L$ in inset of (b-2). In (b), absolute value of $\tn{Im}(kR)$ versus refractive index $n$ is shown, inset (b-1) is an effective potential for $\ell=1$, $m=3,4,5$, and Lamb shift ($L$) versus refractive index ($n$) is also shown in inset (b-2). These two values $|\tn{Im}(kR)|$ and $L$, both of them decreases as $n$ increases and, furthermore, their patterns are very similar to each other.}
\label{Figure-5}
\end{figure}

Let us check these facts in the dielectric cavities. Figure~\ref{Figure-4} shows an effective potential for resonance for $\ell=1$ and $m=3$ with refractive index $n=3.3$ and the boundary $R=1$. Blue dashed line is $k_T^2=9$, solid orange line is $\tn{Re}(kR)^2=\cl{E}\cong 3.42$, and blue dotted line is $k_B^2\cong0.826$. Therefore, the energy of the resonance lies between $k_T^2$ and $k_B^2$, which is so-called trapped resonance. Figure~\ref{Figure-5} compares the eigenvalues of the billiard and dielectric cavity as a function of the refractive index, for fixed radial quantum number $\ell=1$ and angular quantum number ($m=3$, $m=4$, and $m=5$). The range of $n$ is confined from 3.3 to 6. The black colors are for $m=3$, the red ones for $m=4$ and green ones for $m=5$, respectively. Note that the dotted lines are eigenvalues of dielectric cavity, whereas solid lines are those of circular billiard. 

\begin{figure}
\centering
\includegraphics [width=9.4cm]  {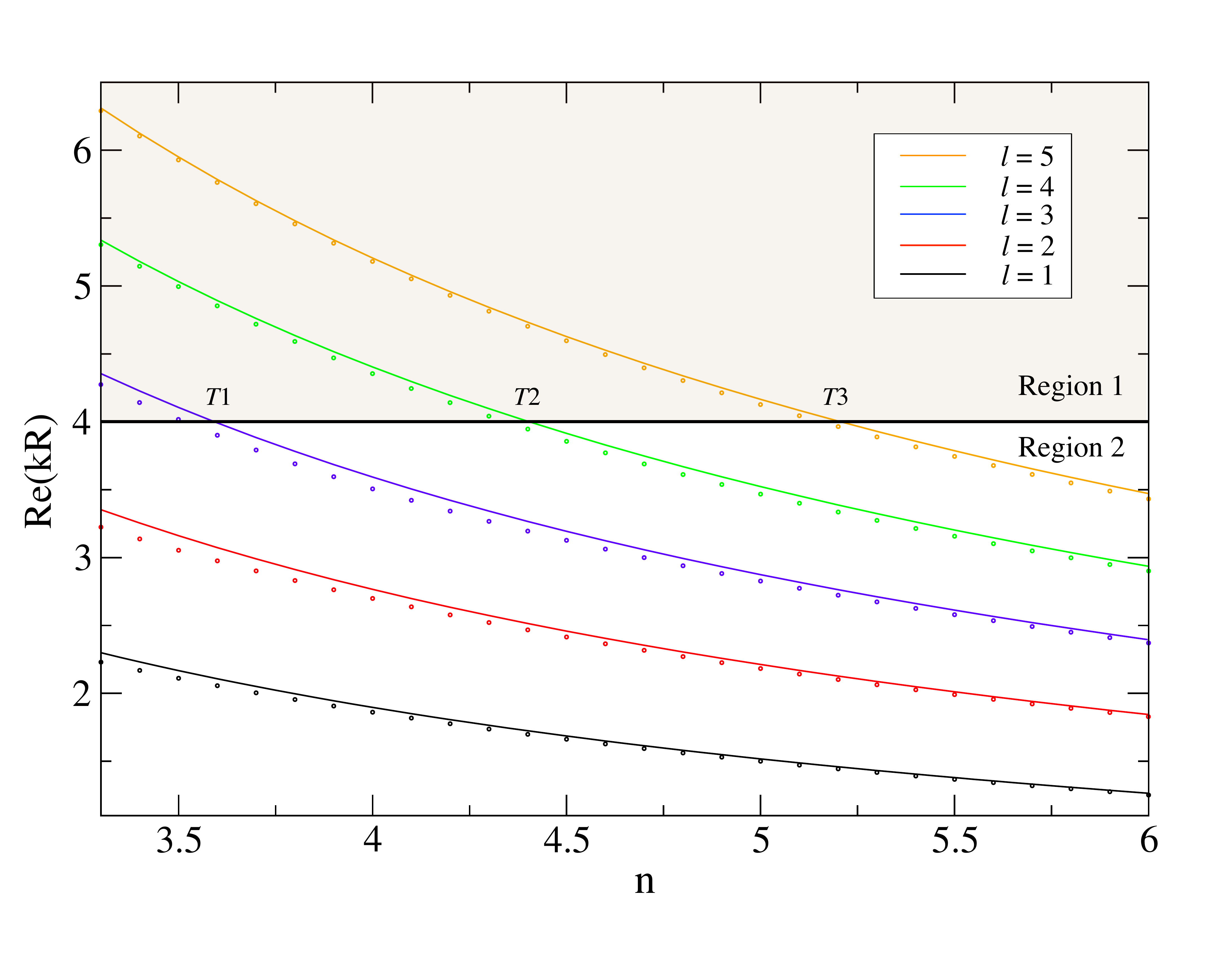}
\caption {(Color online) The eigenvalue trace of $\tn{Re}(kR)$ versus refractive index $n$ with fixed angular quantum number $m=4$ and increasing radial quantum number $\ell=1,2,3,4,5$. The dotted lines are eigenvalues of dielectric cavity whereas solid lines are those of circular billiard. The black colors are for $\ell=1$, red ones are for $\ell=2$, blue ones are for $\ell=3$, green ones are for $\ell=4$, and orange ones are for $\ell=5$. The black thick solid line at $\tn{Re}(kR)$=4 is a top of the trapped region in the case of $R=1$ and $m=4$ by Eq.~(\ref{eq:classicalTP}). This solid line separates Fig.~\ref{Figure-6} into shadow region (Region 1) and (un-shadow) Region 2. In Region 1, the Lamb shift $L$ increases as $n$ increases. On the other hand, it decreases as $n$ increases in Region 2. These are clearly seen in (a) of Fig.~\ref{Figure-8}. The $T1,T2$, and $T2$ points are boundary points of these two regions at $\ell=3$, $\ell=4$, and $\ell=5$, respectively.}
\label{Figure-6}
\end{figure}

The conditions for black dotted curve to be below-barrier resonance is that it must be located between $k_T$ and $k_B$. The $k_T$ is fixed as 3 since $m=3$ and the maximum value of $k_B$ is ${3}/{3.3}\cong0.90$. Therefore, we can easily verify that the black dotted curve entirely lies between $k_B$ and $k_T$ with $[3.3,6]\in n$. Likewise, the other two dotted curves are also below-barrier resonances. 
If we examine the variation patterns of same colored real eigenvalues, we can find out that the difference between dotted and solid lines are getting smaller as $n$ increases. This manifests the decreasing of Lamb shift as $n$ increases as can be seen clearly in inset (b-2) of Fig.~\ref{Figure-5}. Furthermore, the absolute $\tn{Im}(kR)$ also decreases as $n$ increases and their decreasing patterns are very similar to $L$. For this reason, $\tn{Im}(kR)$ is directly related to $L$. This is quite reasonable since we can expect that as $n$ is getting greater, the bottom of potential well ($r=R$) is getting deeper by Eq.~(\ref{eq:effpot}).

Beside, at the constant refractive index $n$, taking as $3.3$, the Lamb shift of $m=5$ is nearly $0.06$, that of $m=4$ is $\simeq0.07$, and of $m=3$ is $\simeq0.085$ for $\ell=1$. In this case, the Lamb shift is getting greater as $m$ decreases, and the bottom of potential well ($r=R$) in Eq.~(\ref{eq:effpot}) is also getting deeper as $m$ increase. (See Fig.~\ref{Figure-5}(b-1).) Consequently, we are apt to conclude that the deeper bottom of potential well lead to a smaller Lamb shift. However, it is not always true. Let us consider the case that we increase $\ell$ instead of $m$, then the situation becomes quite different. There are real eigenvalues of increasing $\ell$ with fixed $m=4$ depending on $n$ in Fig.~\ref{Figure-6}. The black colors are for $\ell=1$, red ones are for $\ell=2$, blue ones are for $\ell=3$, green ones are for $\ell=4$, and orange ones are for $\ell=5$. The dotted lines are eigenvalues of dielectric cavity, whereas solid lines are those of circular billiard, respectively.  

\begin{figure}
\centering
\includegraphics[width=9.2cm]{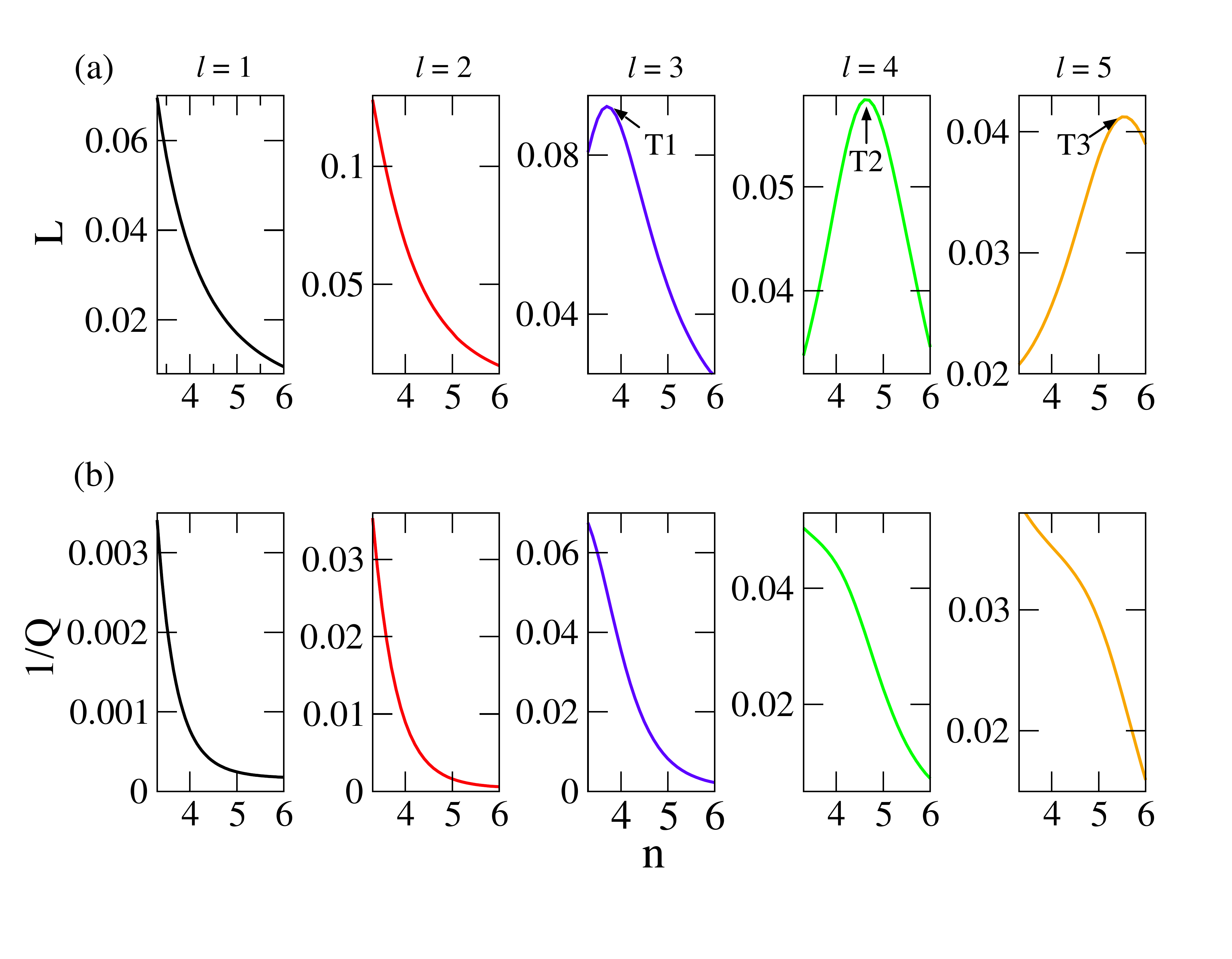}
\caption {(Color online) (a) The Lamb shift $L$ which corresponds to Fig.~\ref{Figure-6}. The black and red curves of $L$ decrease as $n$ increase. Both of them are entirely below $k_T$ in range $3.3\leq n\leq6$. On other hand, the  blue, green, orange curves show special transition patterns whose boundary points are $T1$, $T2$, and $T3$. In blue curve, $L$ increases in $3.3\leq n\leq T1$ and decreases in $ T1\leq n\leq6$. In green and orange colors, $L$ increases in $3.3\leq n\leq T2$ and $3.3\leq n\leq T3$ and decreases in $T2\leq n\leq6$ and $T3\leq n\leq6$, respectively. These boundary points are middle points of between above- and below-barrier. (b) Inverse quality factor versus $n$: The black and red curves below $k_T$ resemble each other very much. But the two pattern are becoming different above $\ell\geq3$.}
\label{Figure-8}
\end{figure}

When we closely examine each colored curve, we can find out that their variation patterns are various. In the case of $\ell=1$ (black) and $\ell=2$ (red), their $L$ are entirely decreasing as $n$ increases, however the other cases are more complicated. 
Here, $L$ first increases but decreases at some threshold values depending on $n$. In order to emphasize these facts, we invoke a black thick solid line at $\tn{Re}(kR)=4$. This solid line separates Fig.~\ref{Figure-6} into shadow region (Region 1) and un-shadow (Region 2). The Region 1 and Region 2 correspond to the \emph{above-barrier resonance} and the \emph{below-barrier resonance}, respectively. The variations of $L$ behave oppositely in each region. In Region 1, the Lamb shift $L$ increases as $n$ increases. On the contrary, it decreases as $n$ increases in Region 2. These individual transitions are clearly seen in Fig.~\ref{Figure-8}(a).
 
 \begin{figure}
\centering
\includegraphics [width=9.2cm] {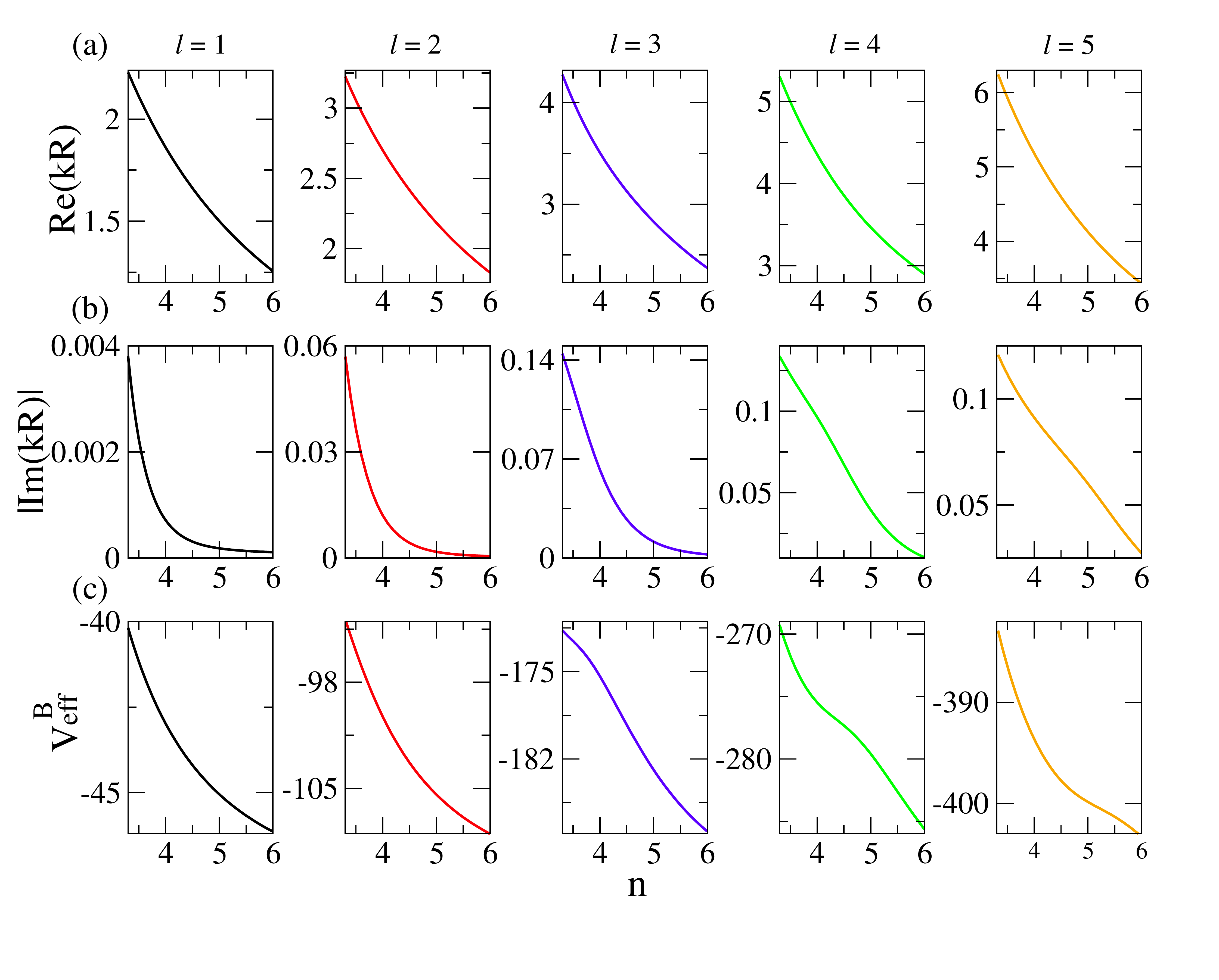}
\caption {(Color online) Trace of $\tn{Re}(kR)$ and $\tn{Im}(kR)$ in Fig.~\ref{Figure-6} simultaneously depend on refractive index $n$. (a) The variation of  $\tn{Re}(kR)$ is getting larger as $\ell$ increases. On the contrary, (b) $\tn{Im}(kR)$ is changing from  exponential decay to almost linear decay, which means that the degree of variation is drastically reduced. (C) The bottom of potential well is getting deeper as $n$ increases.}
\label{Figure-7}
\end{figure}
 
We also adopt threshold points $T$ to show these transitions. These points are boundary points of these two regions at $\ell=3$, $\ell=4$, and $\ell=5$ named $T1$, $T2$, and $T3$, individually. In blue curve, $L$ increases in $3.3\leq n\leq T1$ and decreases $T1\leq n\leq6$. Also, in green and orange colors, $L$ increases in $3.3\leq n\leq T2$ and $3.3\leq n\leq T3$ and decreases in $T2\leq n\leq6$ and $T3\leq n\leq6$, respectively. Therefore, we must need other description to resolve these phenomena. The key is a top of trapped region, i.e., $k_T$=$m/R$. Hence, we usually set $R=1$, it is just 4 on $m=4$, which is a black thick solid line in Fig.~\ref{Figure-6}. The resonances above this top of trapped region is called as above-barrier resonance. All taking together, in the below-barrier resonance, $L$ decreases as $n$ increases. On the other hand, in the above-barrier resonance, $L$ increases as $n$ increases. The clue of this resolution is that the $k_T$ is completely fixed to 4 even though radial quantum number $\ell$'s change.

Thus if we increase $\ell$ with fixed $m=4$, the energy $k^2$ also increases to get away from the potential well and lie on the above $k_T^2$ at some threshold value, which makes un-trapped states. After that, as $n$ is getting larger, the bottom of potential well goes downward respect to the state of energy $k^2$. It means that the scattering state is growing apart from the bottom of potential well. Consequently, the Lamb shift $L$ increases. In our result, the threshold value of $\ell$ is 3 which shows transition at $T1$. In other words, in the above-barrier resonances, when $n$ decreases, the bottom of potential well goes up toward the state of energy $k^2$ to result in diminishing $L$.

In Fig.~\ref{Figure-7}, there are eigenvalues of $\tn{Re}(kR)$, $\tn{Im}(kR)$, and bottom of the potential well $V_{\tn{eff}}^B$ simultaneously depending on refractive index $n$. It shows that $V_{\tn{eff}}^B$ is getting deeper and $|\tn{Im}(kR)|$ decreases, when $n$ increases. The degree of variation of $\tn{Re}(kR)$ is getting larger as $\ell$ increases. On the contrary, in (b), that of $\tn{Im}(kR)$ is changing from exponential decay to almost linear decay, which means that the degree of variation is drastically reduced. These two facts imply that quality factor $Q$ can be also changeable. So, let us investigate (b) of Fig.~\ref{Figure-8}. In order to grasp the relation between $L$ and $Q$, we get the value of inverse quality factor instead of quality factor itself. Thereby, we can get quite similar variation pattern of $L$ and $Q$ in ($\ell=1,2$) with $m=4$ for the below-barrier resonances. This fact is important because the trapped states are usually using lasing mode in experiment. By the way, the two pattern are becoming very much different above $\ell\geq3$. Since this inverse quality factor is wiggling around at the boundary point $T$, it may suggest that the boundary points might be related to the inflection point of inverse quality factor.

\section{Conclusion} \label{conclusion}
We have studied the open nature of dielectric microcavities by comparing the differences  between normal modes and quasi-normal modes (QNMs). The former ones have always real eigenvalues and are defined in solely interior of cavity. On the other hand, The QNMs are defined on the total space comprised of two functions, i.e., $\ket{\phi_k} \tn{and} \ket{\omega_{k}}$. The first one is an eigenfunction of $H_{\tn{eff}}$ whose eigenvalues are generally complex number which is defined entirely on subsystem $S$ and latter one is a resonance tail defined solely on subsystem $E$, and we find out that the real eigenvalues of normal modes are little bit greater than those of the dielectric cavity. 
 
Furthremore, we have investigated the Lamb shift (energy shift due to coupling with the environment) in two circumstances. First, we have found the relation between the angular quantum number $m$ and the Lamb shift $L$. We showed that the Lamb shift decreases as $m$ increases at least in WGMs. Second, we have investigated the relation between the refractive index $n$ and the Lamb shift using effective potential well. There are two kinds of variations in Lamb shifts with respect to $k_T$. In the below-barrier resonances, $L$ decreases as $n$ increases. On the other hand, $L$ increases as $n$ increases in the above-barrier resonances. Moreover, the effective potential well is quite generic in physics so that our methodology might be useful for other fields such as atomic physics and nuclear physics. We hope that our results can give some clues to understanding of the openness effects such as quasi-scar or Goos-H\"{a}nchen shift.

\section{acknowledgment}
We are grateful to Su-Yong Lee and Changsuk Noh for comments.
We thank Korea Institute for Advanced Study for providing computing
resources (KIAS Center for Advanced Computation Linux
Cluster) for this work. This work was partly supported by the IT R\&D program of MOTIE/[10043464]. K.J. acknowledges financial support by the National Research Foundation of Korea (NRF) grant funded by the Korea government (MSIP) (Grant No. 2010-0018295).

\end{document}